\documentclass[reprint,amsmath,amssymb,aps,prb,superscriptaddress]{revtex4-1}
\usepackage{units}
\usepackage{amsmath}
\usepackage{url}
\usepackage{natbib}
\usepackage{textcase}
\usepackage{amssymb}
\usepackage{graphicx}
\usepackage{bm}
\usepackage{float}
\usepackage{multirow,microtype,color,relsize,ulem}
\usepackage{braket}
\usepackage{subfigure}
\usepackage{hyperref}
\usepackage[utf8]{inputenc}
\usepackage[english]{babel}
\hypersetup{colorlinks=true,linkcolor=blue,citecolor=blue}
\hypersetup{linktocpage}

\begin{document}
	
\title{Non-Hermitian disorder in two-dimensional optical lattices}

\author{A. F. Tzortzakakis}
\affiliation{Physics Department, University of Crete, Heraklion, 71003, Greece}
\author{K. G. Makris}
\affiliation{Physics Department, University of Crete, Heraklion, 71003, Greece}
\affiliation{Institute of Electronic Structure and Laser, FORTH, 71110 Heraklion, Crete, Greece}
\author{E. N. Economou}
\affiliation{Physics Department, University of Crete, Heraklion, 71003, Greece}
\affiliation{Institute of Electronic Structure and Laser, FORTH, 71110 Heraklion, Crete, Greece}

\begin{abstract}
	In this paper, we study the properties of two-dimensional lattices in the presence of non-Hermitian disorder. In the context of coupled mode theory, we consider random gain-loss distributions on every waveguide channel (on site disorder). Our work provides a systematic study of the interplay between disorder and non-Hermiticity.  In particular, we study the eigenspectrum in the complex frequency plane and we examine the localization properties of the eigenstates, either by the participation ratio or the level spacing, defined in the complex plane. A modified level distribution function vs disorder seems to fit our computational results. 
\end{abstract}

\date{\today}
\maketitle

\section{Introduction}
\par The study of crystalline solids is based on Bloch’s theorem \cite{Bloch,Blochbook} which assumes a perfect periodicity in the positions of the atoms and in the density of electrons.  However, in actual crystalline solids there are always deviations from periodicity, such as point defects, linear faults (e.g. dislocations), 2D defects (e.g. interfaces of crystallites); if the concentration of these deviations becomes high enough Bloch’s theorem breaks down and a new paradigm emerges featuring novel properties such as the possibility of localized eigenstates \cite{Andersonoriginal,Anderson2D}. The concept of this so-called Anderson localization, which claims that an electronic wave can be trapped in a finite region of a disordered lattice, has been at the center of the attention of the solid state physics community for more than sixty years \cite{50years,Reviewsolids,Scalling,RevNum,Eco3,Pendry}. The localization phenomenon appears due to the interference among multiple scattering processes of the electronic waves by random variations in the potential of the crystal lattice. As a result of this interference, the previously extended eigenmodes of the system, the Bloch waves, may now become localized decaying exponentially for large distances. The phenomenon of Anderson localization has also been studied experimentally, indirectly, by measurements of macroscopic quantities such as the conductance \cite{Mottbook,Lifshitz,Imry,EcoGreen} and the transmission \cite{Landauer,Ecosouk,Datta,ExperimentKramer}. In solid state systems though, the existence of many body interactions and temperature dependent effects, such as inelastic scattering, makes the interpretation of these experiments rather uncertain. In order to overcome this difficulty, the topic of localization was extended to the regime of optics, acoustics and elastics where its consequences were not clouded by other effects producing similar observations \cite{li1,li2,li3,li4,li5,li6,li7,li8}. Such an extension is naturally valid, since the concept of localization  is based on nothing else than wave scattering and interference \cite{Eco2,SoukoulisBook}. The only difficulty with classical waves, such as EM ones, is that they usually exhibit very weak scattering  not enough to produce localization. Several ideas were proposed to circumvent this difficulty, some of which \cite{SJohn,Yablonovitch} led through different paths to photonic crystals \cite{Yablophotonic,Soukoulisphoto,Yablophoto2,Soukoulisphoto2,JoannopoulosBook,SoukoulisBook2} and phononic crystals \cite{SigEco,Revsig,PhononBook,Phonon1,Phonon2,PhonBook,Phonon3}. 
\par In recent years the problem of localization  attracted renewed interest as research moved to two formerly unexplored areas: (a) Many body localization, meaning Anderson localization in the presence of many body interactions \cite{Manybody1,MB2,MB3,MB4} and (b) non-Hermitian systems, such as those obeying parity and time-reversal (PT) symmetry \cite{EP1,EP2,EP3,EP4,EP5}. This recently introduced concept of constructing parity-time (PT) symmetric systems is more appropriate for photonic rather than solid state systems since the former can easily incorporate and realize complex potentials that  require physical gain and loss. Thus PT-symmetry in optics \cite{PT1,PT2,PT3,PT4,PT5,PT6} has been studied extensively over the past few years, leading to the development of the new field of non-Hermitian photonics  \cite{NH1,NH2,NH3,NH4,NH5,NH6,NH7,NH8,NH9,NH10,NH11,ci1,ci2,ci3,ci4,ci5,longhi1}. The possibility for the potential (in our case index of refraction) to possess imaginary part makes the fundamental topic of localization even more complicated and at the same time gives rise to a whole lot of unanswered question about the properties of non-Hermitian disordered systems.
\par In particular, non-Hermitian random matrices are a topic of high research interest in the context of mathematical physics \cite{NH12}, and disordered photonics \cite{rl1,longhi2}. More specifically, random lasers \cite{rl,rl2} where the decay of the cavity modes and the gain material leads naturally to dissipation and amplification, respectively, are a prototypical system in the framework of disordered complex media, where non-Hermiticity plays a crucial role.
\par Apart from random lasers, most works regarding non-Hermitian disorder physics are devoted to Hatano-Nelson matrices \cite{hn}, PT-symmetric random lattices \cite{ptj} and dissipative arrays \cite{rkt}; these works focus mostly on one-dimensional systems with correlated disorder or special symmetries. In this work, we will try to answer some of the main questions that stem from the interplay of disorder and non-Hermiticity in two-spatial dimensions, while comparing our results with the corresponding, well known characteristics of the Hermitian case. In particular, we examine physically realistic Anderson type of non-Hermitian waveguide lattices with the most general uncorrelated disorder that includes gain and/or loss. 
 \par In contrast to the Hatano-Nelson Hamiltonian \cite{hatano1997}, where the non-Hermiticity arises from the off-diagonal elements, due to the applied imaginary vector potential, in our case the non-Hermiticy is a consequence of the complex on-site `energies'. Therefore, the corresponding physical systems, as well as the matrices describing them, are quite different and exhibit different behavior.

\par Our paper is organized as follows: In the next section, Sec. II, we introduce our model possessing diagonal disorder in the real or imaginary or both parts of the potential term $n$; we present also  qualitative data concerning the distribution of eigenfrequencies and the extent of their corresponding eigenfunctions, based on their participation ratio. In Sec. III, we give more quantitative   numerical results  regarding the level spacing/density of states and comment on the comparisons among the three types of disorder. Finally in Sec. IV, we present computational results and comments on the extent of the eigenfunctions in two different ways. 
\section{Non-Hermitian disorder in coupled systems}
We begin our analysis by considering optical wave propagation in a disordered non-Hermitian model, in the context of coupled mode theory \cite{PT4}. We consider a 2D square lattice of $N\times N$ waveguides in the xy plane, with a field propagation constant per waveguide $\{n_{p,q}\}, (p,q=1,...,N)$, which here plays the role of the optical potential. The light propagation along the z axis is described by the following normalized paraxial equation of diffraction:
\begin{equation}
i\frac{\partial \psi_{p,q}}{\partial z}+V(\psi_{p+1,q}+\psi_{p-1,q}+\psi_{p,q+1}+\psi_{p,q-1})+n_{p,q}\psi_{p,q}=0
\label{par}
\end{equation}
where $p,q = 1, 2, ..., N$ , with $N\times N$ being the total number of the waveguides, $\psi_{p,q}$ the modal amplitudes, $V$ the coupling coefficient between two neighboring channels, and $n_{p,q}$  the complex potential strength (field propagation constant) at each waveguide channel. Here we have considered only nearest neighbors interactions and we assume (without loss of generality) that $V=1$. For guided non-Hermitian structures, $n_{p,q}$ is complex and this physically means that each waveguide is characterized by either gain ($Im\{n_{p,q}\}< 0$) or loss ($Im\{n_{p,q}\}> 0$) and by its real part $Re\{n_{p,q}\}$.
\par In order to find the eigenmodes of the system, we substitute $\psi_{p,q}=\phi_{p,q}exp(i\omega_{j} z)$ in the evolution equation (Eq.\ref{par}) and get the eigenvalue problem:
\begin{equation}
\omega_{j} \phi_{p,q}=(\phi_{p+1,q}+\phi_{p-1,q}+\phi_{p,q+1}+\phi_{p,q-1})+n_{p,q}\phi_{p,q}
\label{bas}
\end{equation}
where $\omega_{j}$ is the complex eigenvalue of the $j^{th}$ eigenmode, with $j= 1, 2, ..., N^2$.
\par In a more compact form the above eigenvalue problem can be expressed in terms of a symmetric  tridiagonal matrix $D$ with zeros in the main diagonal and the identity matrix $I$ (both matrices have dimension $N \times N$) by the following relation:
 \begin{equation}
M_{i,j}=z_{i}\cdot\delta_{i,j}+ (D \otimes I + I \otimes D)_{i,j}
\label{matr}
\end{equation}
In the above, $\{z_{i}\}_{i=1}^{N^2}$ is a set of random, complex in general, numbers, located along the main diagonal of $M$. Also, $\otimes$ denotes the Kronecker tensor product between two matrices and the matrix $M$ has dimension $N^2 \times N^2$. 
\par Since $M$ is a non-Hermitian matrix, it is fully described by a set of biorthogonal right $\ket{\phi_{j}^{R}} $ and left $\ket{\phi_{j}^{L}} $ eigenmodes. In other words, we have the following right eigenvalue problem:
 \begin{equation}
M \ket{\phi_{j}^{R}}= \omega_{j} \ket{\phi_{j}^{R}} 
\end{equation}
 and the corresponding left eigenvalue problem of the adjoint matrix:
\begin{equation}
M^{\dagger} \ket{\phi_{j}^{L}}= \omega^{*}_{j} \ket{\phi_{j}^{L}} 
\end{equation}
 The associated biorthogonality condition is $ \braket{\phi_{j}^{L}|\phi_{i}^{R}}=\delta_{i,j}$. In general the right and left eigenvectors are different and, since any dynamics of the problem include both the right and the left set of eigenfunctions, one needs to study both of them (see for example \cite{2rev4,2rev5,2rev6}). In our case though, the left and right eigenfunctions are complex conjugate pairs since $M^{\dagger}=M^{*}$. This is a direct outcome of the corresponding one-dimensional matrices' $I,D$ Hermiticity.
\par So far the discussion was devoted to the general framework of the spatial coupled mode theory that describes the evolution of paraxial waves in two-dimensional optical waveguide lattices. In this paper though, we are interested to investigate the main features of disordered non-Hermitian lattices, and therefore we consider the potential strength $n_{p,q}$ to be a random variable (on-site disorder). Let's start by considering the modal problem. In particular, we are interested in finding how these eigenmodes and their corresponding eigenvalues change when the system becomes disordered. More specifically, we will examine  phenomena related to  Anderson localization in three different cases of disorder: (a) Real disorder, where the potential strength is real: $n=n_{R}$, and possesses a random distribution $ n_{R} \in [-\frac{W}{2},\frac{W}{2}]$, (b) imaginary disorder, where the potential strength is imaginary with a distribution of the form: $n=in_{I}$, $n_{I} \in [-\frac{W}{2},\frac{W}{2}]$ and finally, (c) both real and imaginary disorder, where the potential strength is complex, $n=n_{R}+in_{I}$ with $n_{R} \in [-\frac{W}{2},\frac{W}{2}]$ and $n_{I} \in [-\frac{W}{2},\frac{W}{2}]$. In the above expressions, $W$ is defined as the disorder strength; a uniform distribution within the range $[-\frac{W}{2},\frac{W}{2}]$ in all the cases is assumed. 
\par A value indicative to the localization of the eigenmodes is the participation ratio of each mode ($PR_{j}$), which is given by the relation:
\begin{equation}
PR_{j}=\frac{|\sum_{i=1}^{N^2}|\phi_{j}(i)|^2|^2}{\sum_{i=1}^{N^2}|\phi_{j}(i)|^4}
\end{equation}
where we have used the right eigenvectors and the sum includes all $N^2$ sites of the system. Generally speaking, the PR measures the spread of a state $\ket{\phi}$ over a basis $\{\ket{i}\}_{i=1}^{N^2}$. For weak disorder strength, $PR$ takes values comparable to the system's area, as all lattice sites participate (almost) equally to the eigenfunction. For higher values of $W$, the $PR$ decreases, which means that the eigenmodes tend to become more and more localized. At this point we have to emphasize that the participation ratios based on the left eigenvectors are exactly the same with the above, since $(\ket{\phi_{j}^{L}})^*=\ket{\phi_{j}^{R}}$, according to our previous discussion.
\par We also define the extent length of each mode, which is an easily measured and convenient in some cases quantity, by the relation:
\begin{equation}
\lambda_{j}=\frac{\sqrt{PR_{j}}}{2}
\end{equation}
\begin{figure}[tb]
	\centering
\subfigure{\includegraphics[clip,width=0.5\linewidth]{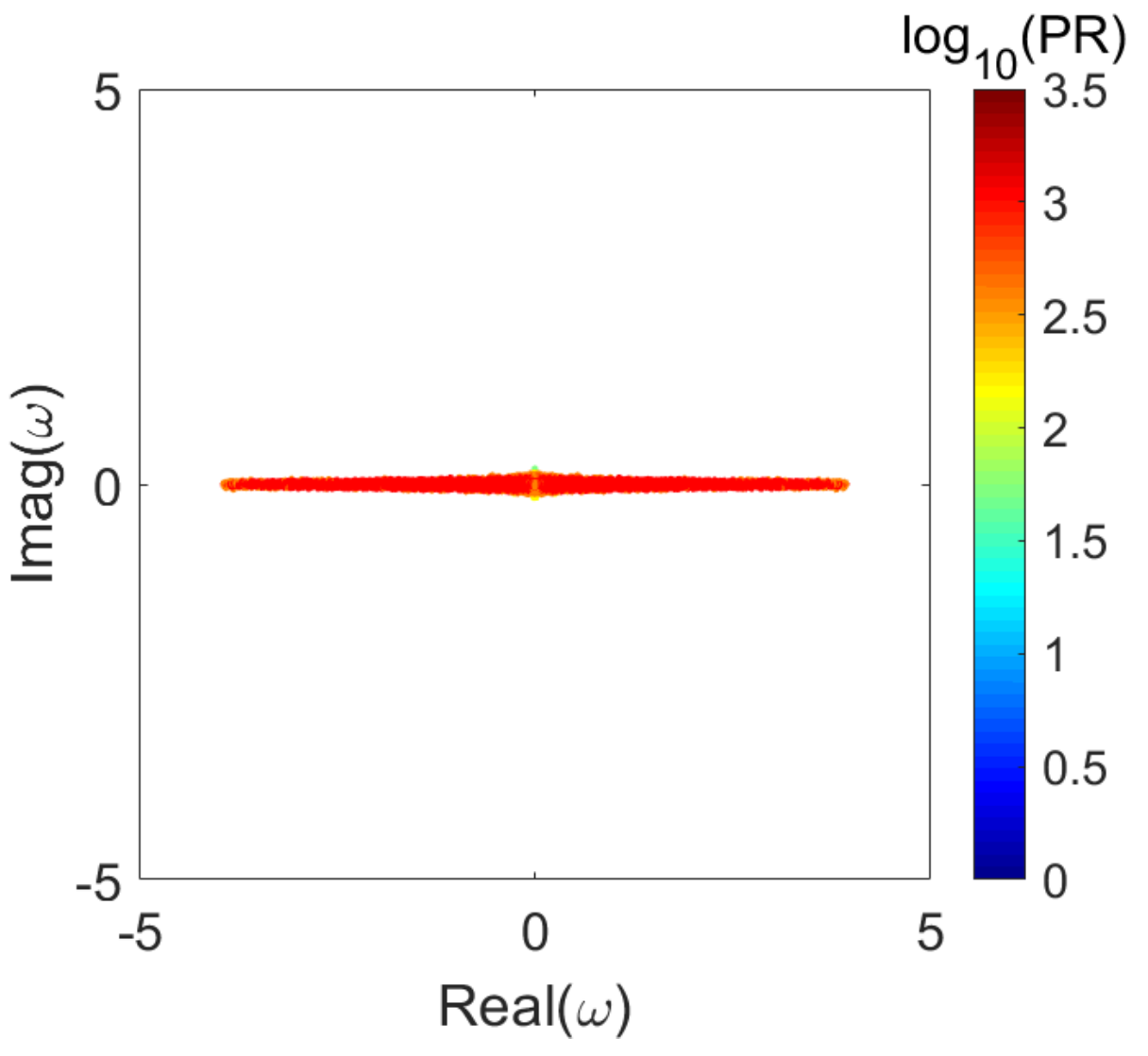}}\subfigure{\includegraphics[clip,width=0.5\linewidth]{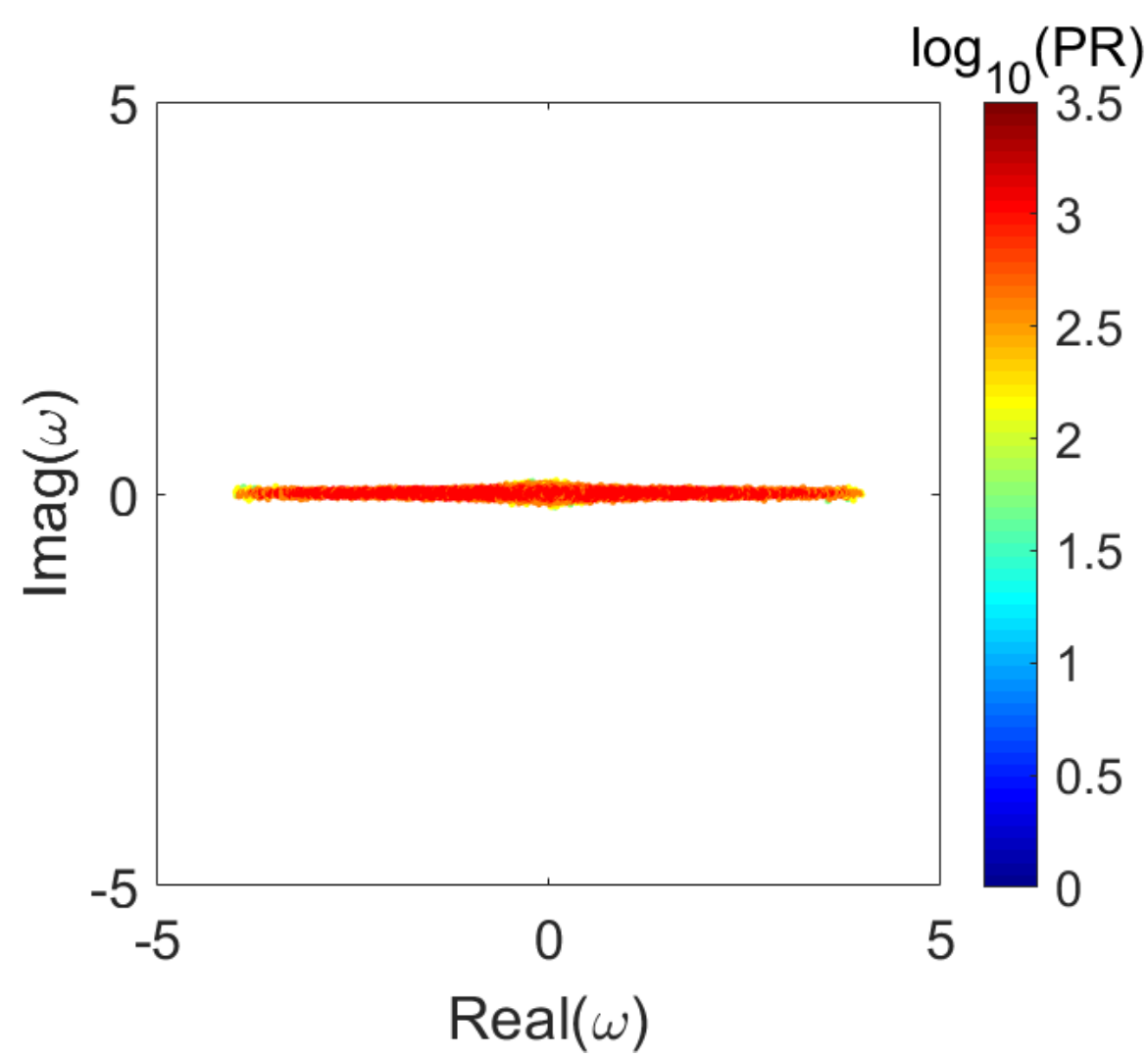}}
(a) W=1
\subfigure{\includegraphics[clip,width=0.5\linewidth]{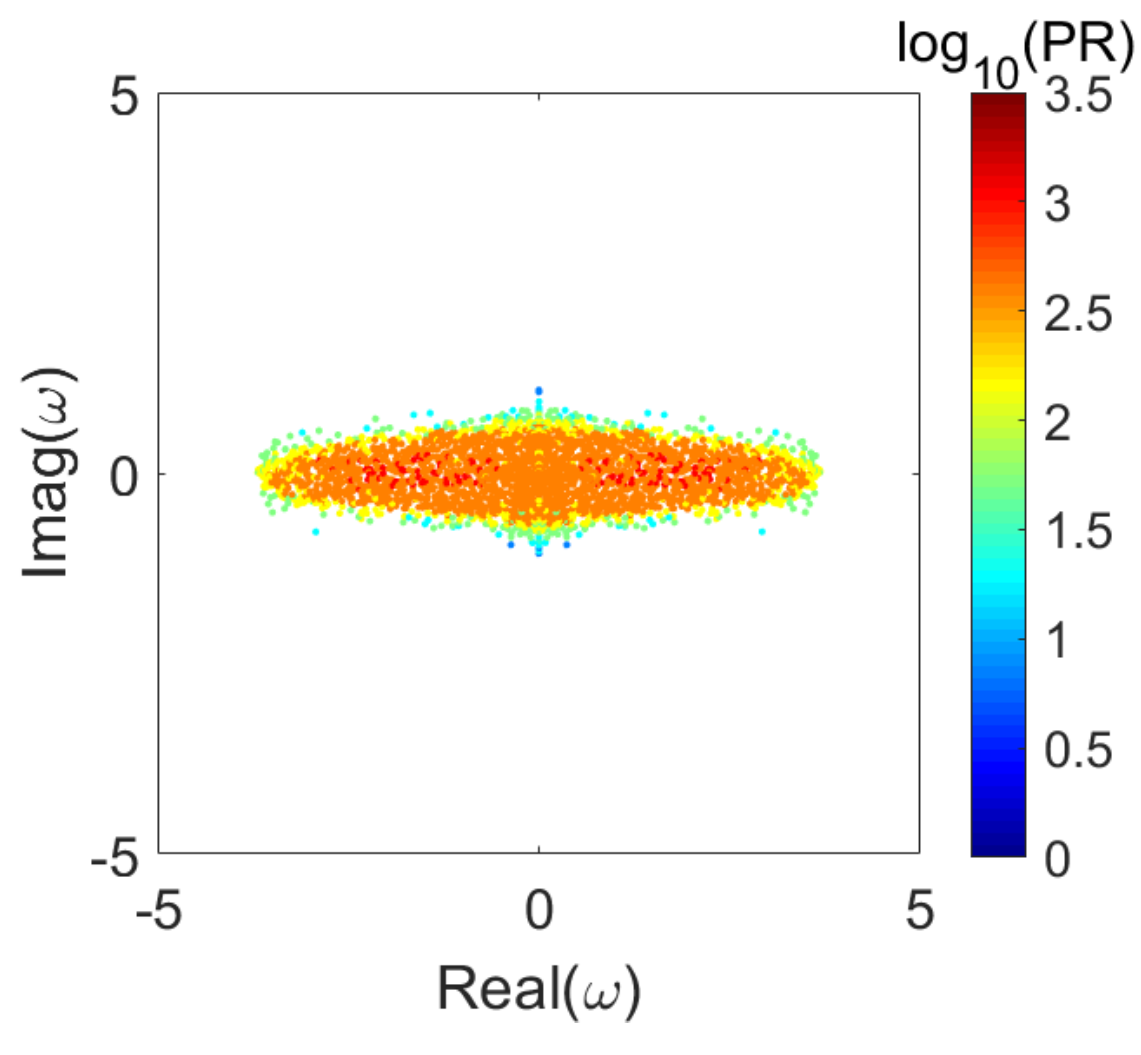}}\subfigure{\includegraphics[clip,width=0.5\linewidth]{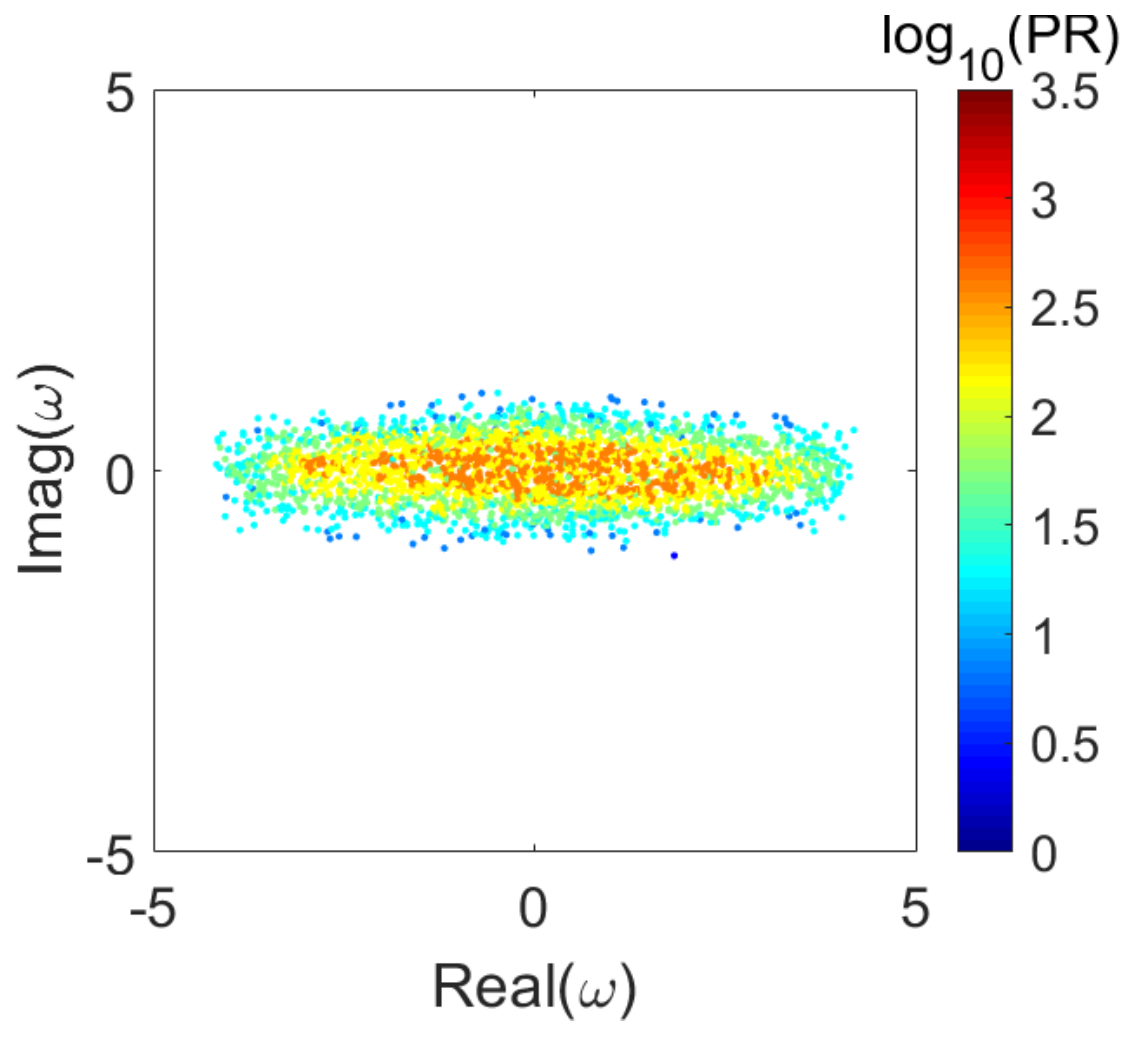}}
(b) W=3
\subfigure{\includegraphics[clip,width=0.5\linewidth]{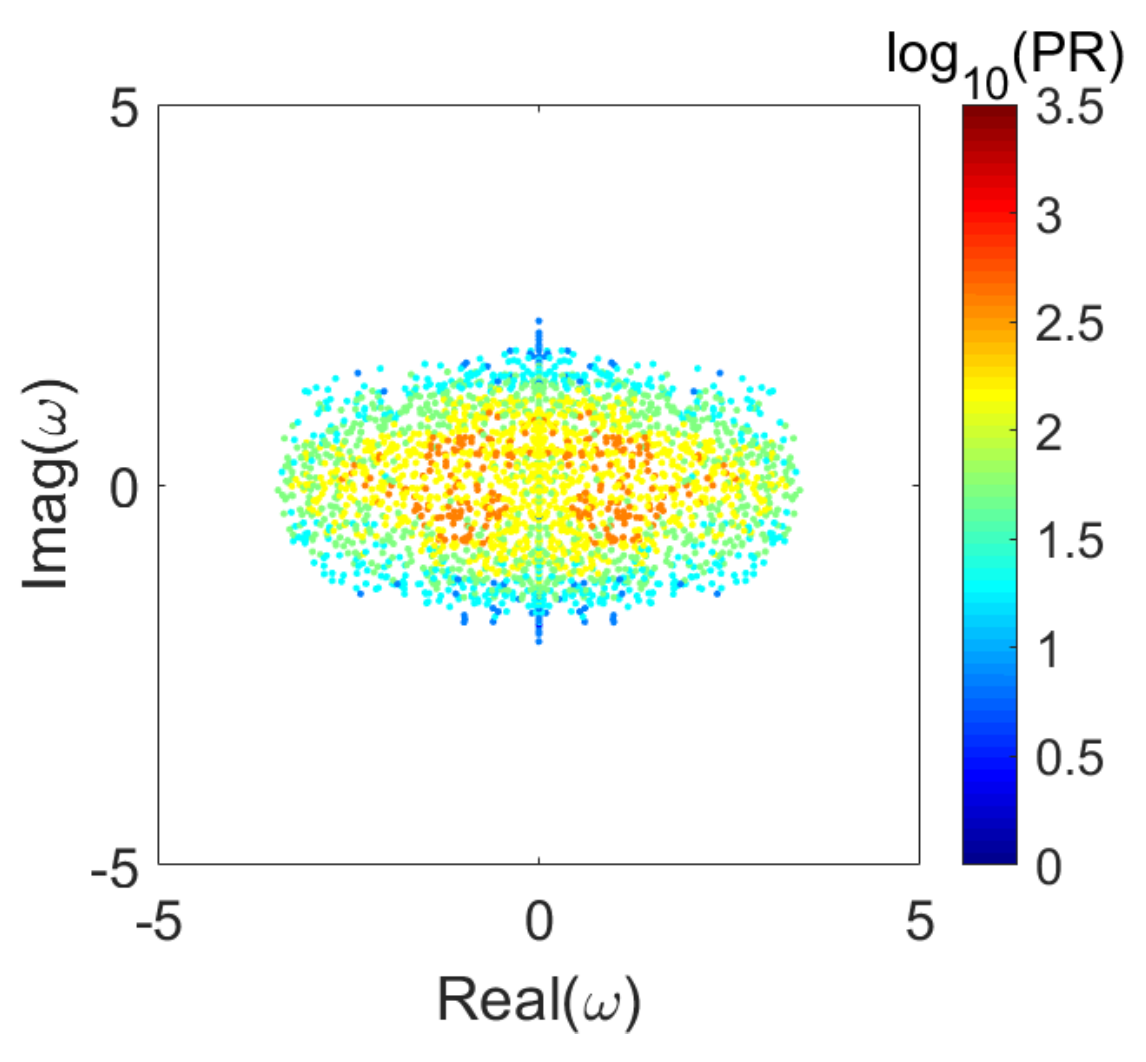}}\subfigure{\includegraphics[clip,width=0.5\linewidth]{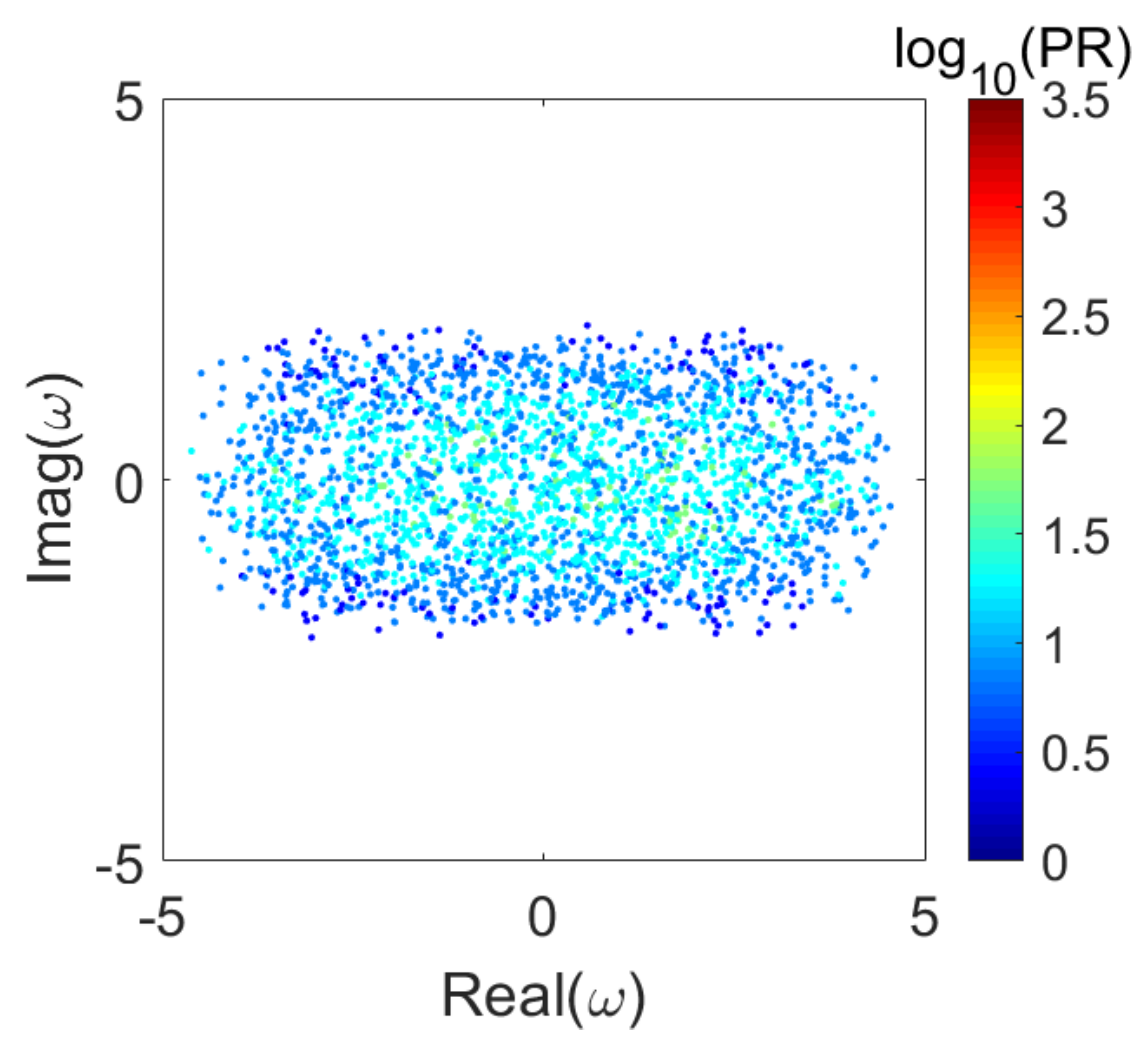}}
(c) W=5
	\caption{Eigenvalue spectra of a 50$\times$50 waveguide lattice (2500 sites) in the complex frequency plane, for a particular realization of the random system and for various disorder strengths. In the left column, disorder is applied only to the imaginary part of the potential strength, while Re($n_{p,q}$)=0; in the right column, disorder of the same $W$ is applied in both the real and the imaginary part of  $n_{p,q}$. Note that the color of each eigenvalue is related to the participation ratio (spatial extension) of the corresponding eigenstate, as seen in the colorbar of the figures.}
	
	\label{eigthree}
\end{figure}
\par To begin with, we will illustrate the eigenvalue spectrum of Eq.\ref{par} in the complex eigenfrequency plane and superimpose the values of $PR$ (of the corresponding eigenmodes) by denoting them with different colors. We do not present any results for the case of real $n$, since they are well known: For weak disorder the density of states ($DOS$) follows more or less the unperturbed $DOS$  (see \cite{EcoGreen}, p.  92) with the rounding of the discontinuities at the band edges and the logarithmic singularity at the band center; states remain essentially extended except at the extreme tails. For strong disorder, the $DOS$ tends to follow the distribution of the potential strength and all states become strongly localized. In Fig.\ref{eigthree}, we show the calculated eigenvalue spectra in the complex frequency plane for various disorder strengths and relate the color of each eigenvalue with the logarithm of the corresponding participation ratio, as seen in the colorbar of the graphs. In this figure we restrict ourselves to only the cases of imaginary disorder (case (b), left column) and the case of disorder in both the real and the imaginary part of the potential (case (c), right column), and for three different disorder strengths. 
\par We can observe that, on both cases, the eigenvalue spectrum forms an approximate ellipse on the complex plane, with a different ellipticity in each case. Note that the spectrum of a system with random diagonal and off-diagonal elements forms a circle on the complex plane. In our case, due to the lack of randomness in the couplings, the obtained spectrum is elliptical \cite{RandomMatrix,Randommatrix3,2rev1,2rev2,2rev3}. As expected, the ellipse is widened for increasing disorder.
\par Most interestingly, we can also observe that, in general, the modes around the center of the elliptical pattern are the most extended ones, while the ones near it's edges are the most localized; this seems to be the non-Hermitian extension of the corresponding well known result from the case of real disorder, in which the eigenstates tend to become gradually localized as we move towards the edges of the bands.

\section{LEVEL SPACING AND DENSITY OF STATES (DOS)}

\par
\begin{figure}[tb]
	\includegraphics[clip,width=1\linewidth]{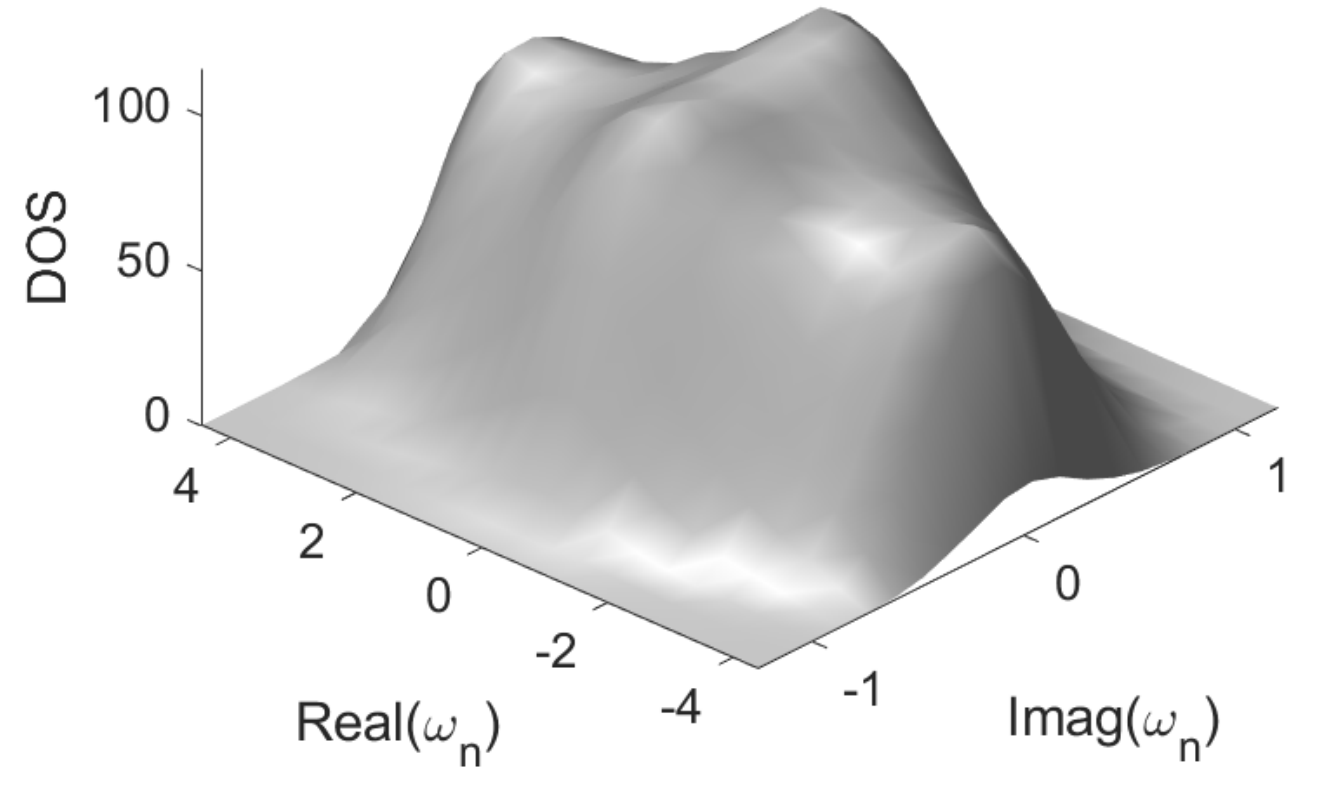}
	\caption{Density of states per unit area in the complex plane Real($\omega$)-Imag($\omega$), for the case of disorder in both the real and the imaginary part of the diagonal matrix elements and disorder strength $W=3$. We have averaged over 1000 realizations of the disorder.}
	\label{Dos}
\end{figure}

 Figure \ref{eigthree} provides a general semi-qualitative picture of the whole spectrum, as well as information about how extended (or localized) the corresponding eigenstates are. However, quantitative information about how dense the spectrum is in each sub-region  of the complex frequency plane cannot be easily inferred from these plots. To remedy this missing information we consider first the density of states  ($DOS$) in the complex frequency plane (averaged over realizations of the random system). To define the $DOS$ we count the number of states $\delta N$ with eigenfrequencies located within an elementary square of area $\delta A=\delta \omega_{R} \times \delta \omega_{I}$ and centered at the point $\omega=\omega_{R}+i\omega_{I}$ of the complex frequency plane; then we have by definition:
\begin{equation}
DOS(\omega)\equiv\frac{\delta N}{\delta A}
\end{equation}
Usually, we implicitly assume that the $DOS$ is averaged over many realizations of the random system.  The $DOS$, as expected,  depends on the disorder strength. For small values of disorder, all the eigenvalues are concentrated near the real axis, while their density is higher in the center and decays as we move towards the edges of the spectrum. On the other hand, for the case of strong disorder, the eigenvalues become almost uniformly distributed in the whole spectrum and tend to follow the same distribution as the diagonal matrix elements, in the limit $W\gg V$, where $V$ is the coupling coefficient (the $DOS$ at the edges drops to zero not discontinuously in contrast to the matrix elements). Most interesting is the case of an intermediate value of disorder, which is shown in Fig.\ref{Dos}, for $W=3$. In this plot we can observe that the $DOS$ shows four weak peaks located on the real and the imaginary axis and near the edges of the spectrum. The peaks are found symmetrically over the center of the complex plane. In addition, the $DOS$ appears to decay as we further move away from these two axis.
\par Regarding the level spacing statistics, there are several ways to define the level spacing, $S$, in the complex frequency plane. One way, termed "1" and for a particular realization of the disorder, is the following:
\begin{equation}
S_{1}(\omega)\equiv\sqrt{\frac{\delta A}{\delta N}}
\label{Spac}
\end{equation}
which is directly related with the $DOS$: $S=(DOS)^{-\frac{1}{2}}$. (This direct relation between $DOS$ and $S$ acquires an extra numerical coefficient if both quantities are averaged over many realizations of the random system).
\par Another way, termed "2", to define the level spacing, $S$, which is the one most commonly used \cite{2rev1,Randommatrix2,MB4}, at each eigenfrequency $\omega_{j}$ is as the minimum distance in the complex frequency plane between two neighboring eigenfrequencies averaged over many realizations of the disorder:
\begin{equation}
S_{2}(\omega)|_{\omega=\omega_{j}}\equiv|\omega_{j}-\omega_{j-1}| 
\label{spacing}
\end{equation}
In the above expression, $\omega_{j-1}$ is the eigenvalue which is nearest to the eigenvalue $\omega_{j}$, on the real axis (Hermitian case) or on the complex plane (non-Hermitian case).
\begin{figure}[tb]
	
	\subfigure[S-Real($\omega$)]{\includegraphics[clip,width=0.5\linewidth]{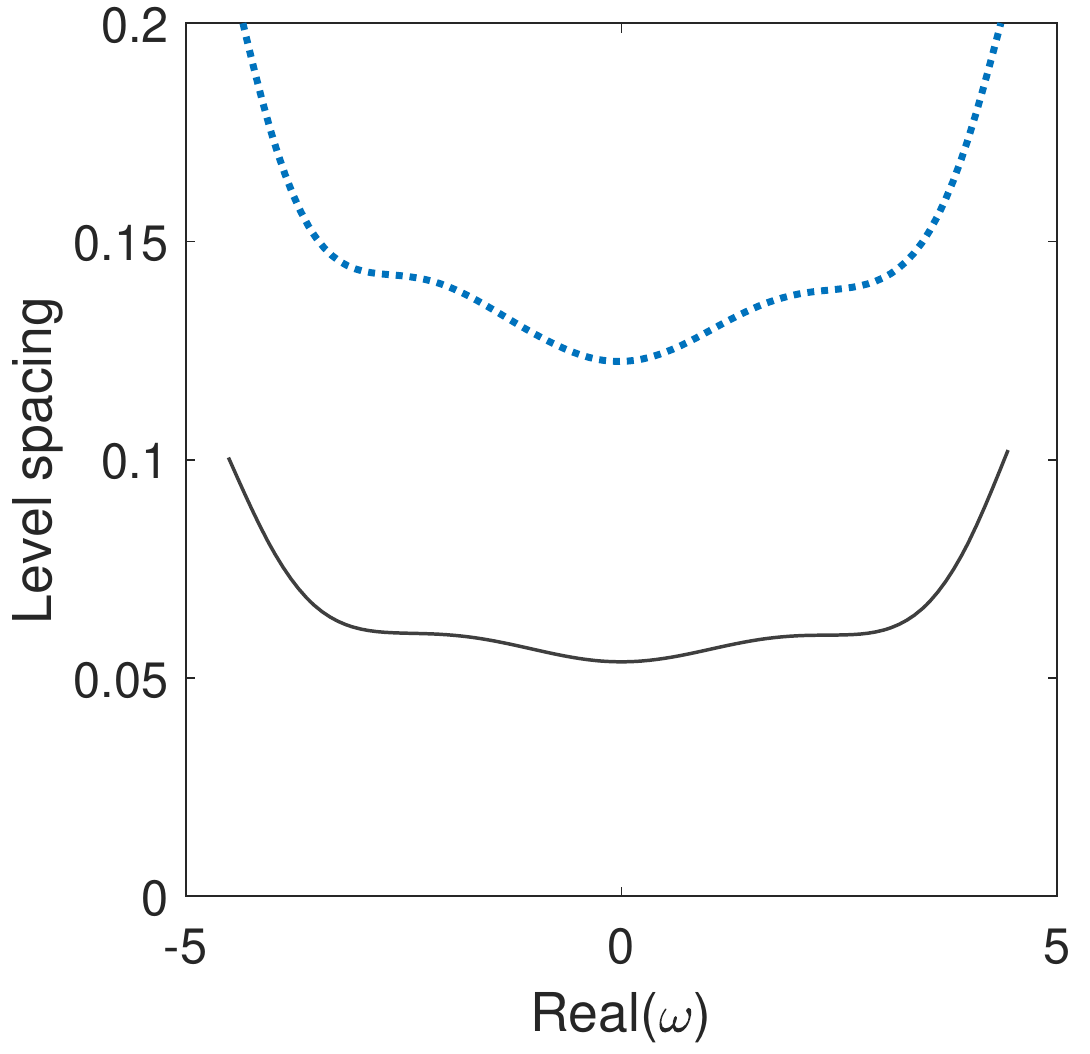}}\subfigure[S-Imag($\omega$)]{\includegraphics[clip,width=0.48\linewidth]{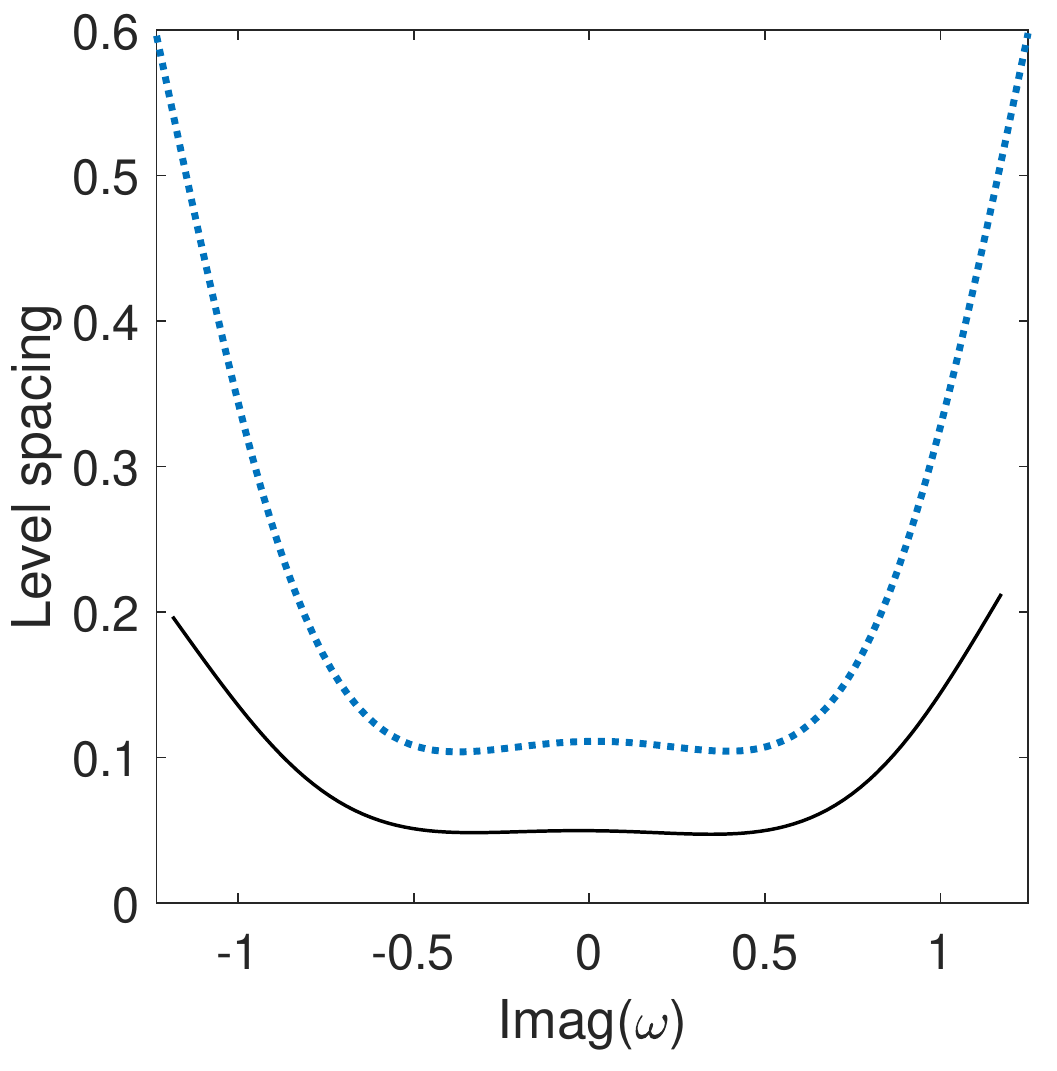}}
	
	\caption{Level spacing $S$ plotted along the real axis (a) and the imaginary axis (b) of the complex frequency plane Real($\omega$)-Imag($\omega$) and for the case of disorder in both the real and the imaginary part of the diagonal matrix elements and disorder strength $W=3$. The blue dotted line represents results of calculations using the definition \ref{Spac} of level spacing ($S_{1}$), while the black line represents results using definition \ref{spacing} ($S_{2}$). An average over 1000 realization of disorder is performed in each case.}
	
	\label{level}
\end{figure}
\par Definition 1 gives the nearest level spacing in the complex frequency plane averaged essentially over all directions in this plane;  definition 2 gives the nearest level spacing along only one direction, the direction which at each eigenfrequency gives the minimum nearest level spacing. It is obvious that the second definition will result systematically in a smaller level spacing, as shown in Fig.\ref{level}. In Fig.\ref{level}(a) we plot the level spacing $S$, according to both definitions 1 and 2, for reasons of comparison, and for $W=3$ in both the real and the imaginary part of the diagonal matrix elements, vs the real frequency axis (i.e. vs $\omega_{R}$ and for $\omega_{I}$=0), while in Fig. \ref{level}(b) along the imaginary frequency axis, (i.e. vs $\omega_{I}$ and for $\omega_{R}$=0). 
 \par We have to note that we deal with non-Hermitian matrices with complex spectra and as a result there is no unique definition of the level spacing, since the second or the third nearest neighbor eigenfrequency can be close to the first in the complex plane. The definition based on Eq.\ref{Spac} is also a measure of the level spacing by being the square root of the area per eigenvalue in the complex plane (essentially averaging over the level spacings within this area); in contrast, the definition based on Eq.\ref{spacing} picks the smaller among these level spacings, and as such is somehow smaller than but similar to that of Eq.\ref{Spac}. Thus the two different measures of level spacing lead to similar, but not identical, conclusions.

\section{SPATIAL EXTENT OF THE EIGENFUNCTIONS}

\par
\begin{figure}[tb]
	\includegraphics[clip,width=1.0\linewidth]{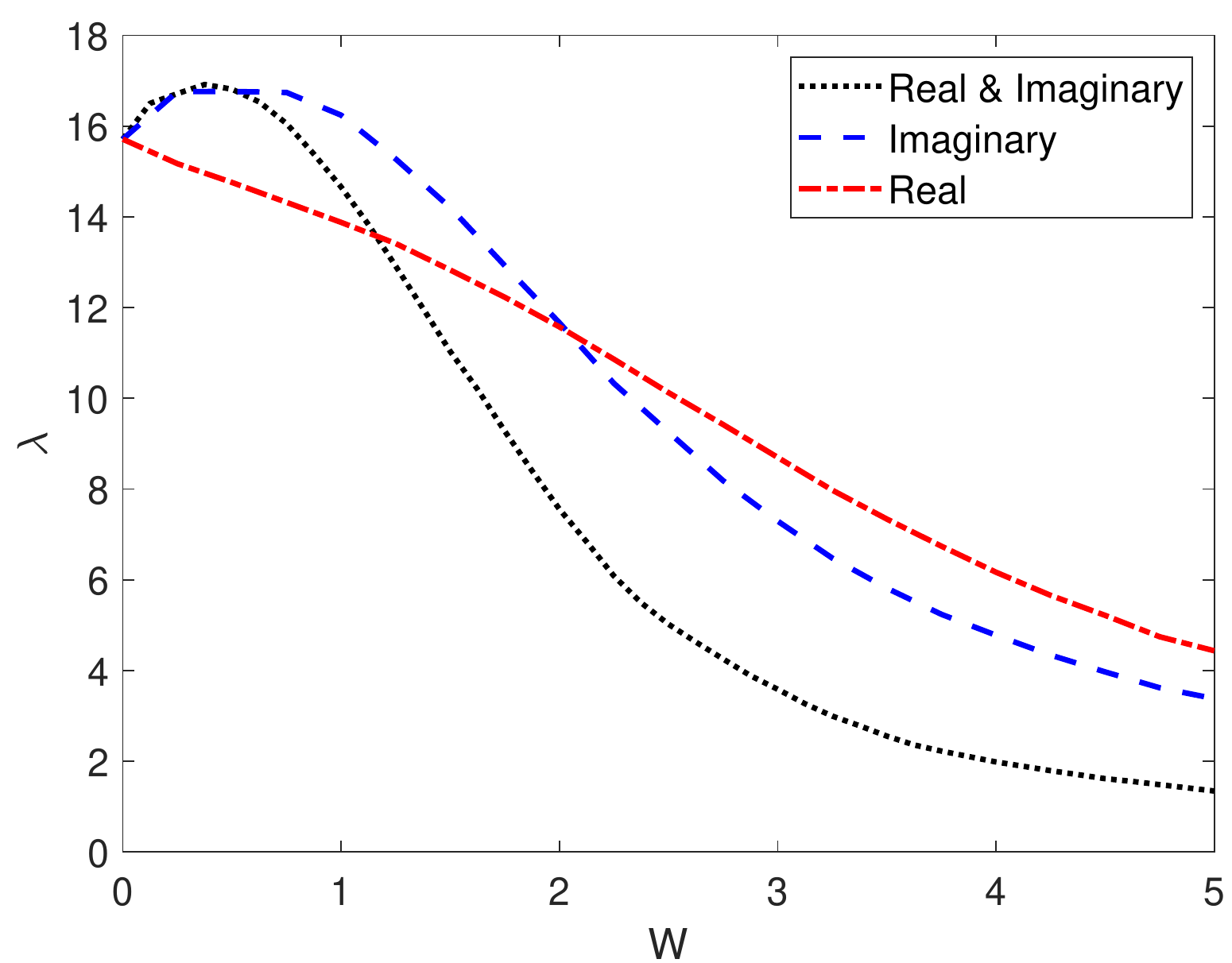}
	\caption{Mean extent length $\lambda$, averaged over the whole spectrum, as a function of the disorder strength W, for a lattice of 50$\times$50 waveguides; uncorrelated disorder for all the three cases: Disorder only in the imaginary part of the diagonal matrix elements (blue dotted line), disorder only in the real part (red dash-dot line) and disorder of the same amplitude in both the real and the imaginary part (black dashed line). }
	\label{ksi}
\end{figure}

 In Fig.\ref{ksi} we present our results concerning the linear extent, $\lambda$, averaged over all the eigenmodes of the system, as a function of the disorder strength: $\lambda=\braket{\lambda_{j}}$. We consider the three different cases:  (1) disorder only in the real part of the diagonal matrix elements; (2) same disorder only in the imaginary, and (3) same disorder in both the real and the imaginary parts. The comparison of the three cases reveals a very interesting feature: While case (1), real disorder, exhibits a monotonic drop of the extent of the eigenfunctions with increasing disorder, as expected, cases (2) and (3), complex disorder, exhibit a surprising increase of the extent of the eigenfunctions with increasing disorder (for small disorder) before they eventually drop even faster than case (1).  We attribute this anomaly for weak disorder to the fact that an imaginary part in the Hamiltonian breaks time reversal symmetry (TRS); it is well known that in Hermitian systems the breaking of TRS (usually by the presence of a static magnetic field) favors delocalization (see \cite{Ecosolid} pp. 510-513). In the present case imaginary disorder acts in a dual way: its implicit TRS breaking favors more extended states, while its disorder nature favors more localized states; the first aspect seems to dominate for only weak disorder, which is not actually surprising: for weak imaginary disorder its breaking of TRS is enough to randomize the phases of closed paths transversed in opposite directions; beyond this point the breaking of TRS has nothing to offer while the further increase of W contributes only to localization.
\par The distribution of the level spacing averaged over the whole spectrum provides in the  Hermitian case valid information \cite{Randommatrix2} about the localization or not of the eigenfunctions. This  distribution of nearest-neighbor level spacing  often takes one of several universal forms, depending only on the disorder strength and not on the details of the medium \cite{Levelstat,Randommatrix4}. More specifically, when it comes to wave propagation in disordered media (either classical or quantum), this distribution is related with the spatial extent of the system's modes. 

\begin{figure}[tb]
	\includegraphics[clip,width=1\linewidth]{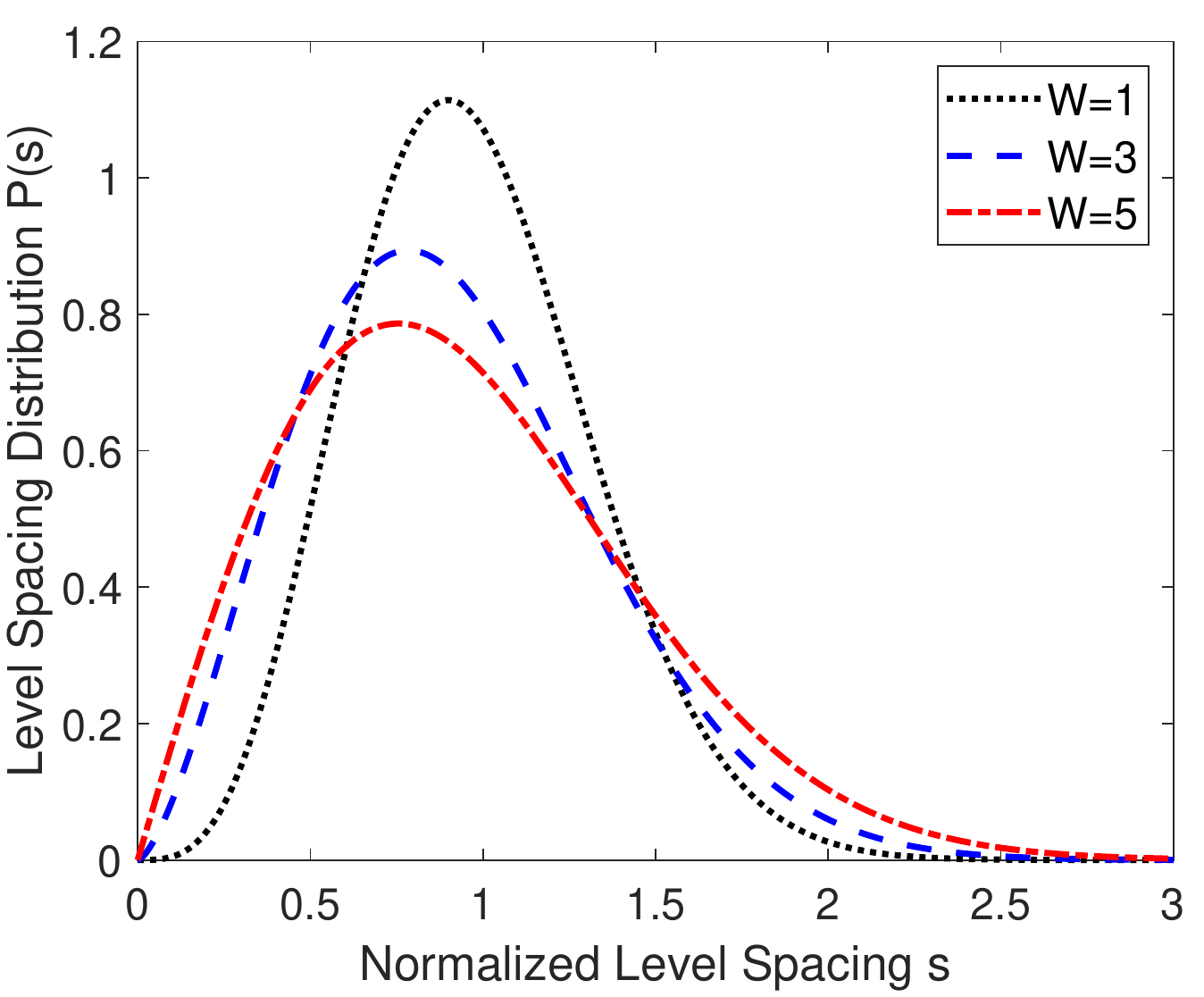}
	\caption{Probability density P of normalized level spacings S, as defined by Eq.\ref{spacing}, averaging over the whole spectrum. In this plot, disorder is applied in both the real and the imaginary part of the potential. The results for three values of disorder strength are shown: $W=1$ (black dotted line), $W=3$ (blue dashed line) and $W=5$ (red dash-dot line). An ensemble of 50 realizations of disorder is used in each case.}
	\label{lvl}
\end{figure}
\par In the Hermitian case \cite{Randommatrix4,Randommatrixbook}, for weak disorder, the normalized eigenvalue spacings obey the Wigner-Dyson distribution: $P_{WD}(S)=aS\cdot exp(-bS^2)$. This is a direct consequence of the extended character of the modes which implies their strong overlap and, consequently,  the so-called level repulsion which does not allow two eigenvalues to almost coincide: $P_{WD}(S=0)=0$. 
\par On the other hand, in the regime of strong disorder, the eigenvalue spacings obey the Poisson distribution: $P_{P}(S)=exp(-S)$. Here, due to the strong localization of the eigenmodes, there is very little  spatial overlap and, as a result, there is no level repulsion and, hence, their eigenvalues can be arbitrarily close to each other.

\begin{figure}[tb]
	\subfigure[]{\includegraphics[clip,width=0.5\linewidth]{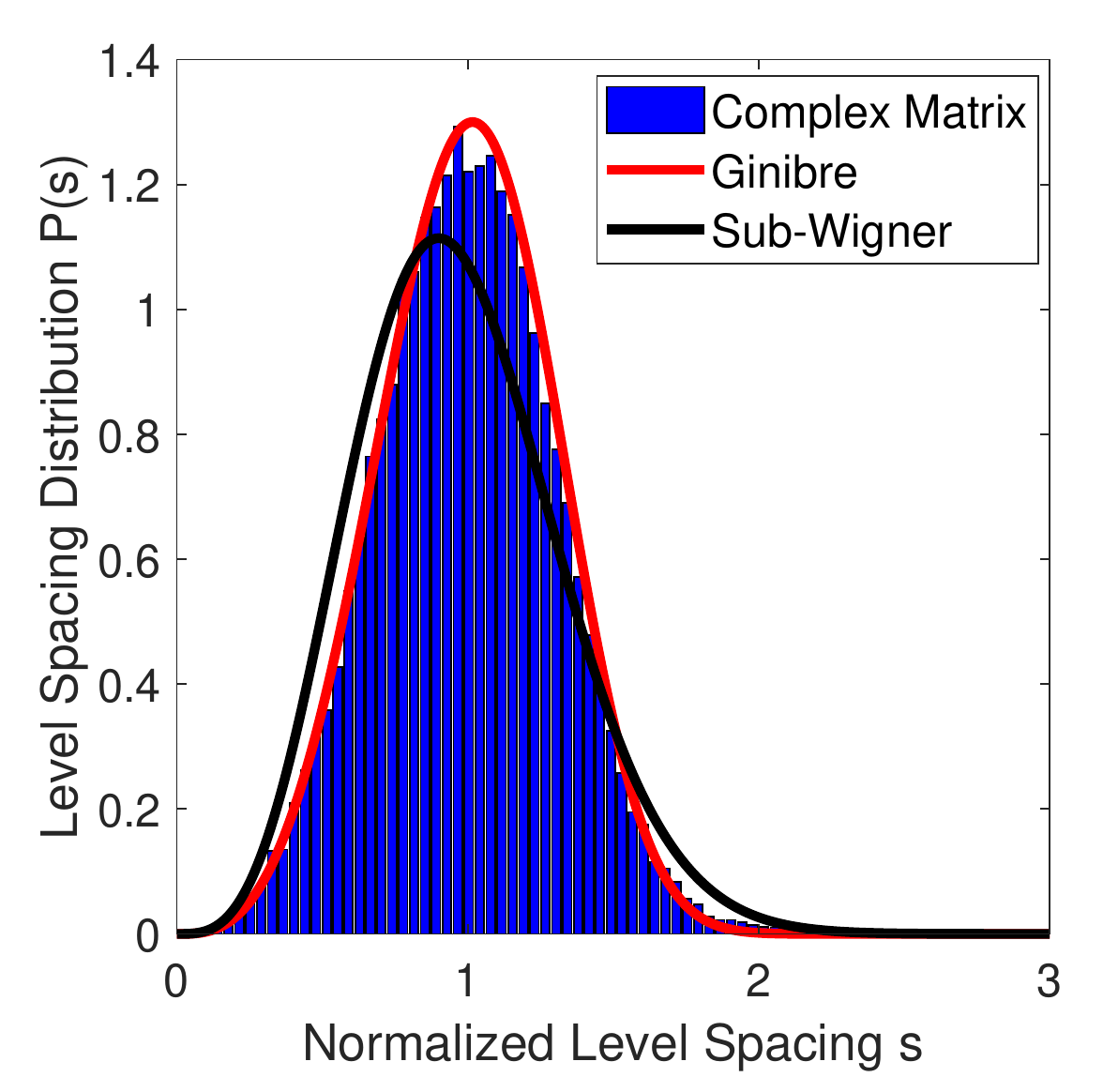}}\subfigure[]{\includegraphics[clip,width=0.5\linewidth]{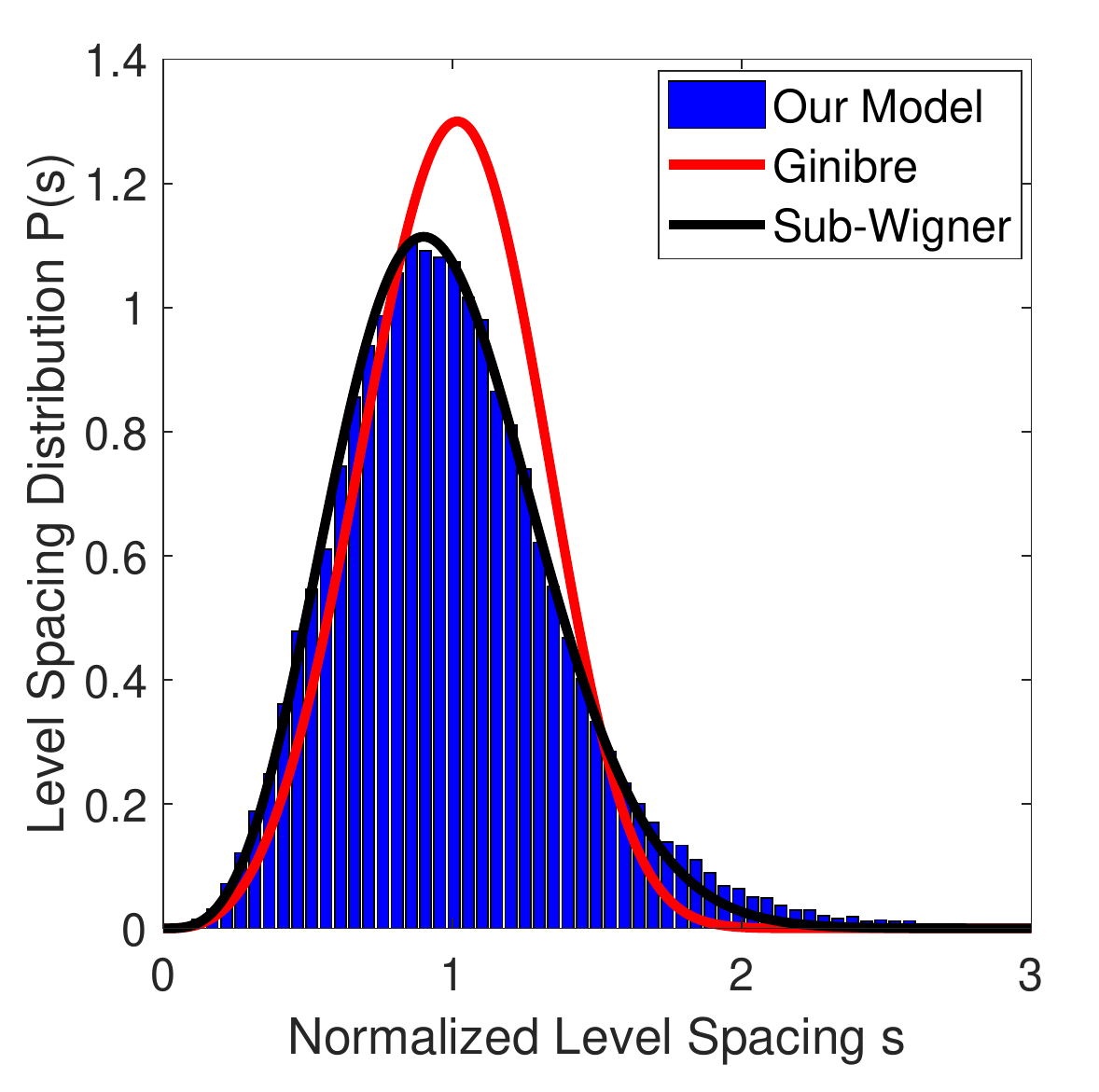}}
	\caption{Probability densities $P$ of normalized level spacings $S$, as defined in Eq.\ref{spacing}, averaged over the whole spectrum for disorder strength $W=1$ and for the same size. (a) $P$ for a full complex matrix with random elements (blue) compared with Ginibre (red) and sub-Wigner (black). (b) $P$ for our model (blue) compared with Ginibre (red) and sub-Wigner (black). An ensemble of 50 realizations of disorder is used in each case.}
	\label{distr}
\end{figure}
\par When disorder is applied on the imaginary part of the potential though, the eigenvalue spacings (according to definition 2, Eq.\ref{spacing}) appear to obey a different probability distributions than the ones referred above. In figure \ref{lvl} we show the level spacing distribution for the case of  both real and imaginary disorder and for three different values of disorder strength ($W=1$, $W=3$ and $W=5$). The results for  only imaginary disorder do not differ significantly from the ones shown in the figure. We can observe that, for the above cases, the level spacing distribution can be described from a sub-Wigner ($SW$) probability distribution curve fitting:
\begin{equation}
	P_{SW}(S)=aS^b\cdot exp(-cS^2), \quad b \geq 1
\end{equation}
For $W=1$, the exponent is $b\simeq 3$ and as we raise the disorder this exponent is gradually lowered (for $W=3$, $b\simeq 1.5$). For $W=5$, $b=1$ and the level spacing distribution obeys the Wigner-Dyson probability distribution $P_{WD}(S)$. Thus, we find that the level repulsion undergoes a smooth transition from $S^3$ to $S^1$ with increasing disorder; a similar behavior to the one found in random matrix theory papers, such as \cite{2rev7}.
\par In the strong disorder regime ($W=5$ in Fig.\ref{lvl}, we find that $P(S)=P_{WD}(S)$. Since our level statistics take place in the 2-D complex plane, this distribution expresses the Poissonian statistics of uncorrelated eigenvalues located in 2 spatial dimensions \cite{MB4,grobe1988quantum,markum1999non}.
\par On the other hand, in the case of weak disorder ($W=1$), the level spacing distribution is a sub-Wigner one which exhibits cubic level repulsion ($P(S)\sim S^3$ as $S\to 0$). Since this dependence is similar to the corresponding dependence of the Ginibre distribution \cite{Randommatrix2,ginibre,cubrep}, one may conclude that our result is nothing more than the Ginibre distribution. As we show in Fig.\ref{distr}, that is not true. In order to compare our distribution with the Ginibre we calcluate the corresponding distribution for a full complex matrix and compare with our result of Fig.\ref{lvl} (black dotted line). The conclusion of such comparison, based on Fig.\ref{distr}, is clear. The level spacing statistics of our matrix is not properly described by the Ginibre distribution, as our model's matrix is a sparse (block tridiagonal) matrix which possesses the additional symmetry: $M^{\dagger}=M^{*}$. These two features make the level spacing statistics different from the well known Ginibre distribution, which coincides with the the full complex matrix and not with the sub-Wigner (Fig.\ref{distr}(a)); in contrast our distribution is clearly different from the Ginibre and coincides with the sub-Wigner (Fig.\ref{distr}(b)).
\section{Discussion}
Before concluding, we would like to comment on a crucial point: is there an Anderson transition in a 2D non-hermitian system, or all states are localized even for very week disorder as in the Hermitian case? As we pointed out in the previous paragraphs, the presence of weak imaginary disorder, by breaking the time reversal, reduces the probability of a quantum particle to remain in the same region and hence it favors delocalization. This tendency is further supported by the level spacing exhibiting for weak disorder an $S^{3}$ dependence as $S$ tends to zero;  moreover, the increase of the extent length shown in Fig.\ref{ksi} for weak increasing imaginary disorder provides further support to the tendency for delocalization in the presence of weak imaginary disorder. At this point we think that is difficult to definitely conclude if there is an Anderson transition in non-hermitian 2D case. Nevertheless, in a recent preprint based on our work \cite{nhand}, Huang and Shklovskii reached the conclusion that the localization properties in our model are qualitatively the same as in the corresponding Hermitian model, namely that all states are localized in 2D and that there is an Anderson transition in 3D; this transition occurs at substantially lower critical disorder than that of the Hermitian case ($W=6.15$ vs $W=16.5$). The authors based their conclusion on the numerical behavior of the ratio of the second to the first nearest neighbor level spacing (see \cite{huse} for a detailed analysis of the method), both defined as in Eq.\ref{spacing}.

\section{Conclusions}

\par In conclusion, we have examined the localization phenomenon of the eigenmodes of two-dimensional random optical lattices, in the presence of non-Hermitian diagonal disorder. We have found that the spectrum of such a system forms an approximate ellipse on the complex plane, with the eigenvalues located near the middle of the ellipse to correspond to the less localized eigenfunctions. In addition, the breaking of time reversal symmetry in the non-Hermitian case, favors delocalization for weak disorder, while, as disorder is increased, the localization induced by the non-Hermitian disorder appears to be even stronger than in the Hermitian case. Finally, the level spacing distribution, averaging over the whole spectrum, seems to obey a sub-Wigner probability distribution, when non-Hermitian disorder is applied. For greater values of disorder, the corresponding distribution turns to obey the Wigner-Dyson distribution, which in fact represents the Poissonian statistics in the complex plane.

\section{Acknowledjments}
E. N.  Economou acknowledges  support  by  the European Research Council under ERC Advanced Grant no. 320081 (project PHOTOMETA) and the European Union’s Horizon 2020 Future Emerging Technologies call (FETOPEN-RIA) under grant agreement no. 736876 (project VISORSURF).
\medskip

\bibliographystyle{apsrev}
\bibliography{sample}

\end{document}